\documentclass[aps,groupedaddress,showpacs,showkeys,twocolumn,
longbibliography]{revtex4-1}

\usepackage{xcolor}
\usepackage{graphicx}
\usepackage{url}
\usepackage{amsmath}
\usepackage{amsfonts}
\usepackage{color}
\usepackage[utf8]{inputenc}

\newcommand{\bs}[1]{\boldsymbol{#1}}

\begin{document}
 
\title{Bayesian Approach for Linear Optics Correction}
 
\author{Yongjun Li} \thanks{Email: yli@bnl.gov}
\author{Robert Rainer}
\author{Weixing Cheng}

\affiliation{Brookhaven National Laboratory, Upton, New York 11973}
 
 
\begin{abstract}
With a Bayesian approach, the linear optics correction algorithm for
storage rings is revisited.  Starting from the Bayes' theorem, a complete
linear optics model is simplified as ``likelihood functions'' and ``prior
probability distributions''.  Under some assumptions, the least square
algorithm and then the Jacobian matrix approach can be re-derived.  The
coherence of the correction algorithm is ensured through specifying a
self-consistent regularization coefficient to prevent overfitting.
Optimal weights for different correction objectives are obtained based on
their measurement noise level. A new technique has been developed to
resolve degenerated quadrupole errors when observed at a few select BPMs.
A necessary condition of being distinguishable is that their optics
response vectors seen at these specific BPMs should be near-orthogonal.
\end{abstract}
 
\maketitle

\section{\label{sect:intro}introduction}
At modern particle accelerator facilities, advanced beam diagnostics
instruments with high acquisition rate can generate copious amounts of
data within a short time period. The data can be used in a wide variety of
applications such as characterizing machine parameters, monitoring machine
performance, realizing real-time correction, feedbacks, etc.  A specific
example would be obtaining beam turn-by-turn (TbT) data from beam position
monitors (BPM) after the beam is disturbed. The linear optics functions,
such as, the envelope function $\beta$ of betatron oscillation and its
phase $\phi$~\cite{Courant:1997rq}, can be
extracted~\cite{Castro:1996,Irwin:1999,Huang:2005}.  Due to various
measurement noise, accurately identifying quadrupole error sources is
important for optics correction.  One can average over repetitive
measurements, then use the mean values directly.  Distributions of
measurement noise, which are usually ignored, however, can provide rich
information for identifying error sources precisely. Using a Bayesian
approach and the information provided by the error analysis, the linear
optics correction problem presented by accelerators can be approached from
the viewpoint of probability.

Lattice measurement noise and quadrupole excitations errors are usually
randomly distributed around their expectation values.  Overfitting
quadrupole errors must be avoided.  Specifically, the optics functions
$\beta$ and $\phi$ can be measured with BPMs at many locations $s_i$,
where $i=0,1,\cdots,N-1$, and $N$ is the total number of BPMs. Given a set
of measured data with noise, fitting the actual quadrupole errors $\Delta
K$, is a typical nonlinear regression problem since the dependence of
$\beta$ and $\phi$ on $K$ is nonlinear.  In regression problems,
overfitting is a modeling error which occurs when a function is too
closely fit to a limited set of data points~\cite{Bishop:2006,Yaser:2012}.
There are two reasons of revisiting this problem with a Bayesian approach.
First, the Bayesian approach is a proven technique in preventing
overfitting.  Second, several optics distortions caused by quadrupole
errors need to be corrected simultaneously, but measured in different
units and scales.  With the Bayesian approach, the coherence of the
correction algorithm, which is capable of dealing with multi-objective
regression problems, can be established.

In some scenarios, an optics distortion pattern is indeed caused by a
single quadrupole error rather than normally distributed errors.  However,
the goal of the Bayesian approach is to distribute the error to multiple
sources.  It can sometimes fail to distinguish the single source from its
highly degenerated neighbors.  A new technique has been developed where
only a few specific BPMs are selected to address the degeneracy.  One
necessary condition for being distinguishable is that the optics response
vectors of those specific BPMs should be near-orthogonal.

To further explain this approach, the remaining sections are outlined as
follows: Sect.~\ref{sect:rm} briefly review the well-known correction
scheme, i.e., use of Jacobian matrix as well as a discussion of the
difficulties of this method.  In Sect.~\ref{sect:probability}, we start
from the Bayesian theorem to re-derive the least square algorithm. It will
become clear that using Jacobian matrix is a simplified version of
Bayesian approach with a flat prior probability.  Sect.~\ref{sect:deg}
introduces a new technique for resolving the degeneracy of neighboring
quadrupoles.  Both simulation and experimental data taken from the
National Synchrotron Light Source II (NSLS-II) storage ring are used to
illustrate this technique. A brief summary is given in
Sect.~\ref{sect:summary}.

\section{\label{sect:rm}Jacobian matrix approach}
A linear optics model of a storage ring can be represented by a set of
$s$-dependent optics functions. For simplicity's sake, the envelope
function $\beta(s)$ is used as an example. Other lattice functions such as
betatron phase, $\phi(s)$ will be covered later.  Given a fixed magnetic
lattice layout, an optics model reads as
\begin{equation}\label{eq:beta}
\beta = \beta(s,\bs{K}),
\end{equation}
here $\bs{K}$ is a vector composed of all normalized quadrupole focusing
strengths, and $s$ is the longitudinal coordinate. Bold symbols, such as
``$\bs{X}$'', are used to denote vectors and matrices throughout this
paper. The design lattice model is represented as
\begin{equation}\label{eq:beta0}
\beta_{0}(s) = \beta(s,\bs{K}_0),
\end{equation}
where $\bs{K}_0$ represents the quadrupoles' nominal setting, $\beta_0$
is the nominal envelope function along $s$. If quadrupoles have some
errors $\Delta\bs{K}$ on top of $\bs{K}_0$, the linear lattice is
distorted as
\begin{equation}\label{eq:beta-beat}
\beta=\beta_{0} +\Delta\beta=\beta(s,\bs{K}_0+\Delta\bs{K}).
\end{equation}
$\Delta\beta$ is often referred to as $\beta$-beat.

Given the total distribution of $N$ BPMs, a ring's optics model can be
simply represented as an $N$-dimensional vector.  Once the linear optics
is measured at these BPMs, the lattice correction needs to identify
(account for) the actual quadrupole errors.  A well-known and
straightforward correction method is to use the linear dependence of the
$\beta$-function on each quadrupole, i.e., its Jacobian
matrix~\cite{Huang:2005},
\begin{equation}\label{eq:dbdk}
  M_{i,j} = \frac{\partial\beta_{s_i}}{\partial K_j},
\end{equation}
where $s_i$ is the $i^{th}$ BPM's longitudinal location, and $j$ is the
index of quadrupoles. $\bs{M}$ can be constructed based on either a design
model or a beam-based measurement.  By solving the following equation
\begin{equation}\label{eq:response}
 \Delta\bs{\beta} = \bs{M}\cdot\Delta\bs{K},
\end{equation}
the error sources $\Delta\bs{K}$ can be identified approximately.  Here
$\Delta\bs{\beta}$ is a vector of the $\beta$-beat seen at the BPMs. Since
the dependence isn't linear, iteratively applying Eq.~\eqref{eq:response}
is needed.

It turns out that, due to measurement noise, highly degenerated
quadrupoles and even bad BPMs, the solution to Eq.~\eqref{eq:response} is
spoiled by overfitting.  It is well-known that overfitting can be
prevented by a regularization
technique~\cite{Bishop:2006,Yaser:2012,Huang:2007zzr}. But how to choose a
reasonable regularization coefficient to cut off measurement noise isn't
obvious here.  At the same time, multiple optics functions are measured,
but in different scales and units.  We can stack their response matrices
vertically with some weights.  However, the strategy for specifying
appropriate weights isn't clear.  An inappropriate regularization and
weight specification could degrade the performance of correction.  For
example, some functions are overfitted and can sacrifice others.  Another
concern is that the quadrupole errors obtained with
Eq.~\eqref{eq:response} usually reproduce the optics distortions at the
location of the BPMs rather than the whole ring.  This might be a critical
issue for collider rings, in which no BPM can sit exactly at their
interaction points (IP).  Minimizing the $\beta$-beat at IP's neighboring
BPMs doesn't ensure that the optics at IPs are optimized.  Therefore, it
is important to precisely identify errors in order to correct lattice
distortion globally, rather than limited to the locations of BPMs.  Some
of these difficulties can be mitigated by using a complete accelerator
model instead of its Jacobian matrix.  Another method is to validate the
obtained quadrupole errors at a few testing BPMs, which are intentionally
left out from the fitting.  However, iterative fitting and validation with
a complete optics model is time-consuming, especially for large scale
storage rings.  In the next section, the lattice correction problem will
be addressed using the Bayesian approach.

\section{\label{sect:probability}Bayesian approach}
From the viewpoint of probability, identifying quadrupole errors from
repetitive and independent measurements can be achieved by computing a
posterior conditional probability distribution and determining its maxima.
Consider a simple case of $\beta$ function in the horizontal plane.  Based
on the Bayes' theorem, the conditional probability of having an error
$\Delta\bs{K}$ with a measured $\bs{\beta}$ reads
as~\cite{Bishop:2006,Bernardo}
\begin{eqnarray}\label{eq:bayes}
  p(\Delta\bs{K}|\bs{\beta}) & = &
  \frac{p(\bs{\beta}|\Delta\bs{K})
    p(\Delta\bs{K})}{p(\bs{\beta})} \nonumber \\ &
  \propto &
  p(\bs{\beta}|\Delta\bs{K})p(\Delta\bs{K}).
\end{eqnarray}
Eq.~\eqref{eq:bayes} can be interpreted as, given a measured optics
distortion $\bs{\beta}=\bs{\beta_0}+\Delta\bs{\beta}$, the probability of
it being the error source of $\Delta\bs{K}$ is proportional to the product
of a likelihood function $p(\bs{\beta}|\Delta\bs{K})$ and a probability
distribution of error $\Delta\bs{K}$.  The likelihood function can be
recognized as being related to the dependence of $\beta$ on $K$, i.e., the
Jacobian matrix of Eq.~\eqref{eq:dbdk}. $p(\Delta\bs{K})$ is known as
prior probability distribution which will be covered in greater detail
later.  $p(\bs{\beta})$ is the normalizing constant.

By maximizing the probability in Eq.~\eqref{eq:bayes}, the most likely
quadrupole error distribution can be obtained.  In general, we can assume
that both $\beta$ measurement noise and quadrupole excitation errors are
normally distributed.  For example, at a particular BPM, repetitive
measurement of $\beta$s are distributed around an expectation value
$\mathbb{E}(\beta)=\bar{\beta}$ with a variance $\sigma_{\beta}$.
\begin{equation}\label{eq:normdist}
  \mathcal{N}(\beta|\bar{\beta},\sigma_{\beta}^2)=\frac{1}{\sqrt{2\pi}\sigma_{\beta}}
  \exp\left[-\frac{(\beta-\bar{\beta})^2}{2\sigma_{\beta}^2}\right].
\end{equation}
Eq.~\eqref{eq:bayes} thus can be re-written as
\begin{equation}\label{eq:normal}
 p(\Delta\bs{K}|\bs{\beta}) \propto
 \mathcal{N}(\bs{\beta}|\bs{\bar{\beta}}(s_i,\Delta\bs{K}),\sigma^2_{\beta})
 \cdot\mathcal{N}(\Delta{\bs{K}}|\bs{K}_0,\sigma^2_K).
\end{equation}
Where $\sigma_K$ is the variance of quadrupole error distribution.
Maximizing the probability of Eq.~\eqref{eq:normal} is equivalent to
minimizing its negative logarithm,
\begin{eqnarray}\label{eq:lstsq}
 -\ln\left[p(\Delta\bs{K}|\bs{\beta})\right]& \propto &
 \frac{1}{2\sigma^2_{\beta}}\sum_i
 \left[\beta(s_i,\Delta\bs{K})-\bar{\beta}(s_i)\right]^2 +
 \nonumber \\
 & & \frac{1}{2\sigma^2_K}\Vert\Delta\bs{K}\Vert^2.
\end{eqnarray}
Here $\Vert\bullet\Vert$ is the Euclidean norm of a vector.
Eq.~\eqref{eq:lstsq} can be recognized as the least-square algorithm but
with some well-defined weights.  It is important to note that, since we
assume a normal distribution for quadrupole errors, the solution to
Eq.~\eqref{eq:lstsq} is intended to allocate errors according to a normal
distribution, even if they are not.  If there is a systematic calibration
error on the quadrupole excitations, the second distribution in
Eq.~\eqref{eq:normal} has an non-zero mean. But after the first several
iterations, the non-zero mean value should be able to be filtered out.  A
more detailed discussion on a single outlier of quadrupole error will be
addressed in Sect.~\ref{sect:deg}.

Now we take a look at the first term on the right-hand side of
Eq.~\eqref{eq:lstsq}. By expanding $\beta$ with respect to quadrupole
errors $\Delta{\bs{K}}$ at $\beta_0$ and keeping the linear components, it
reads as
\begin{eqnarray}\label{eq:linearization}
& &\frac{1}{2\sigma^2_{\beta}}
\sum_i
\left[\beta_0(s_i)+\frac{\partial\beta(s_i)}{\partial\bs{K}}\Delta{\bs{K}}
  -\bar{\beta}(s_i)\right]^2
\nonumber \\
&=&\frac{1}{2\sigma^2_{\beta}}
\sum_i
\left[\frac{\partial\beta(s_i)}{\partial\bs{K}}\Delta{\bs{K}}
  -\Delta\bar{\beta}(s_i)\right]^2,
\end{eqnarray}
where $\Delta\bar{\beta}(s_i)=\bar{\beta}_0(s_i)-\beta_0(s_i)$. After
differentiating every term with respect to $\Delta\bs{K}$ and expressing
it in the format of a matrix, we obtain the well-known
Eq.~\eqref{eq:response}. The solution to Eq.~\eqref{eq:response} is given
as
\begin{equation}\label{eq:inverse}
\Delta\bs{K}=\left[\bs{M}^T\bs{M}\right]^{-1}\bs{M}^T\Delta\bs{\bar{\beta}}.
\end{equation}
$\left[\bs{M}^T\bs{M}\right]^{-1}\bs{M}^T$ is often known as the pseudo
inverse of $\bs{M}$, because $\bs{M}$ is usually non-invertible.

\begin{figure}[!ht]
\centering \includegraphics[width=1.\columnwidth]{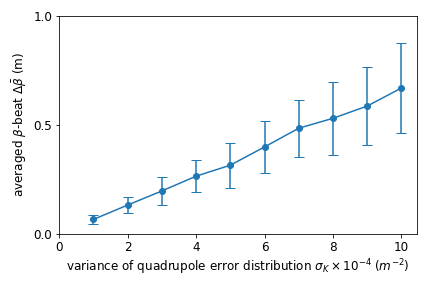}
\caption{\label{fig:dbxVsQuad} Statistic illustration of the horizontal
  $\beta$-beat due to the random quadrupole errors. This linear
  correlation with gradually increasing variance are calculated with the
  NSLS-II ring lattice model in advance. Once an averaged $\beta$-beat is
  measured, its corresponding variance of quadrupole error distribution
  can be used as the prior probability to prevent overfitting}
\end{figure}

Thus far, the measurement noise $\sigma_\beta$ has been ignored. The
solution to Eq.~\eqref{eq:inverse} often overfits quadrupole errors from
either noisy BPM data, or even bad BPMs if they are present.  The
overfitting can be mitigated by taking the second term into account, which
is known as regularization technique.  By adding an additional penalty
term to the sum of squares in Eq.~\eqref{eq:lstsq}, one can prevent the
fitted quadrupole errors from deviating from a reasonable normal
distribution.  In other words, a complete linear optics model provides
not only a likelihood function but an informative prior probability
distribution of quadrupole errors as well.  The solution to the
least-squares problem with regularization is
\begin{equation}\label{eq:regularization}
  \Delta\bs{K}=\left[\bs{M}^T\bs{M}+\lambda\bs{I}\right]^{-1}
  \bs{M}^T\Delta\bs{\bar{\beta}}
\end{equation}
It is important to note that the optimal regularization coefficient
$\lambda=\frac{\sigma^2_{\beta}}{\sigma^2_K}$ is well-defined here.  More
specifically, the variance $\sigma_k(\Delta\bar{\beta})$ of the quadrupole
error distribution $p(\Delta\bs{K})$ should be determined by the measured
$\beta$-beat level using the designed lattice model.
Fig.~\ref{fig:dbxVsQuad} illustrates that the horizontal $\beta$-beat is
linearly proportional to the variance of quadrupole error distribution at
the NSLS-II ring.  After averaging repetitive $\beta_x$ measurements and
comparing against the nominal $\beta_{x,0}$, the variance of quadrupole
error probability distribution $\sigma_k(\Delta\bar{\beta})$ can be
determined with Fig.~\ref{fig:dbxVsQuad}.  During an iterative correction,
$\beta$-beat reduces gradually, as do the corresponding quadrupole errors.
Therefore the regularization coefficient should be dynamically adjusted to
speed up the convergence as well.  The $p(\Delta\bs{K})$ is named as the
prior probability because it can be estimated
analytically~\cite{Courant:1997rq} or numerically in advance.  In the
previous section, one can still use the regularization technique to avoid
overfitting, but the coefficient is not necessarily optimal due to lack of
a theoretical basis.  Experimentally one can obtain this regularization
coefficient based on correction performance on a trial
basis~\cite{Huang:2007zzr}.  However, the Bayesian approach can explicitly
give its statistic and physics interpretation.

Now consider the case of having multiple correction objectives. With
precisely aligned TbT data, the betatron oscillation envelope function
$\beta$ and its phase $\phi$ can be obtained simultaneously.  The strategy
for balancing the correction among multiple objectives is straightforward
in the Bayesian approach.  The probability of having an error
$\Delta\bs{K}$ with given measured $\bs{\beta}$ and $\bs{\phi}$ is a
product of two conditional probability distributions
\begin{eqnarray}\label{eq:beta_phi}
  p(\Delta\bs{K}|\bs{\beta},\bs{\phi}) & \approx & 
  \frac{p(\bs{\beta}|\Delta\bs{K})p(\bs{\phi}|\Delta\bs{K})
    p(\Delta\bs{K})}{p(\bs{\beta},\bs{\phi})} \nonumber \\
  & \propto & 
  p(\bs{\beta}|\Delta\bs{K})p(\bs{\phi}|\Delta\bs{K})p(\Delta\bs{K}).
\end{eqnarray}
Maximizing the probability yields
\begin{eqnarray}\label{eq:lstsq2}
 -\ln\left[p(\Delta\bs{K}|\bs{\beta},\bs{\phi})\right] & \propto &
 \frac{1}{2\sigma^2_{\beta}}\sum_i
 \left[\bs{\beta}(s_i,\Delta\bs{K})-\bs{\bar{\beta}}(s_i)\right]^2 +
 \nonumber \\
 & &\frac{1}{2\sigma^2_{\phi}}\sum_i
 \left[\bs{\phi}(s_i,\Delta\bs{K})-\bs{\bar{\phi}}(s_i)\right]^2 +
 \nonumber \\
 & & \frac{1}{2\sigma^2_{K}}\Vert\Delta\bs{K}\Vert^2.
\end{eqnarray}
After minimizing and linearizing the first two sums of squares in
Eq.~\eqref{eq:lstsq2}, two response matrices with different weighted
blocks can be vertically stacked as
\begin{equation}\label{eq:weighted_matrix}
\bs{\mathcal{M}} = 
\left[
\begin{matrix}
 \bs{M}_{\beta}\\
\frac{\sigma^2_{\beta}}{\sigma^2_{\phi}}\bs{M}_{\phi}\\
\end{matrix}
\right],
\end{equation}
where the weight for $\beta$-block has been normalized to 1. The
significance of the weight coefficient
$\frac{\sigma^2_{\beta}}{\sigma^2_{\phi}}$ of $\phi$-block is to allow for
correcting $\beta$ and $\phi$ coherently.  In other words, no objectives
are over-emphasized by sacrificing others.  The objective, which can be
measured more precisely (with a smaller variance), plays more important
role in the process of correction, automatically, as it should. The
overfitting of Eq.~\eqref{eq:weighted_matrix} can be mitigated with the
same regularization technique as Eq.~\eqref{eq:regularization}, in which
$\bs{M}$ needs to be replaced by the stacked matrix $\bs{\mathcal{M}}$.
The product of two conditional probability distributions can be extended
to cover multiple distributions, for example, $\beta$, $\phi$ and
dispersion $\eta$ on two separate planes.

\begin{figure}[!ht]
\centering \includegraphics[width=1.\columnwidth]{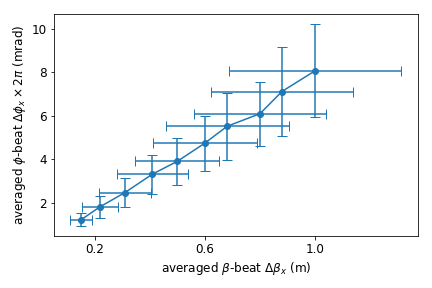}
\caption{\label{fig:betaVsPhi} The correlation of $\beta_x$ and $\phi_x$
  distortion distribution for the NSLS-II ring. This correlated
  distribution is calculated with the lattice model and could be used as
  another prior probability to eliminate incoherent measurement data. The
  farther away from the line, the less reliable the measured data is. Once
  a data point is not within the range of the error bars, it should be
  removed from the data pool.}
\end{figure}

Strictly speaking, in Eq.~\eqref{eq:beta_phi}, the probability
distributions, $p(\bs{\beta},\bs{\phi}|\Delta\bs{K})$ can be expressed as
the products of $p(\bs{\beta}|\Delta\bs{K})$ and
$p(\bs{\phi}|\Delta\bs{K})$ only when they are completely independent.  In
reality they are correlated by
\begin{equation}\label{eq:phi}
\phi(s_1\rightarrow s_2) = \int_{s_1}^{s_2}\frac{1}{\beta(s)}\mathrm{d}s.
\end{equation}
Numerically we can use a lattice model to compute the correlation
distribution of $\Delta\beta$ and $\Delta\phi$ for given quadrupole error
distributions.  Both their mean and variance have a linear dependence on
quadrupole errors as shown in Fig.~\ref{fig:betaVsPhi} for the NSLS-II
ring.  If measured data are within this range they can be treated
approximately as two independent probability distributions.  If not, the
measurement data is not reliable, and should be discarded.

Both $\beta$ and $\phi$-functions at different locations have different
sensitivities to the quadrupole errors~\cite{Courant:1997rq}.  With a
designed lattice model, one can compute another prior probability to
specify different weights for each term in Eq.~\eqref{eq:lstsq} and
\eqref{eq:lstsq2} to further optimize the algorithm.  This process might
be necessary for collider rings because the variation of $\beta$ is
extremely large around interaction points, i.e., the final focus sections.

\section{\label{sect:deg}resolving degeneracy}
In the previous section, we discussed the case in which the lattice
distortion is due to normally distributed quadrupole errors. Once a real
error is localized in a particular quadrupole, it may require us to
identify which quadrupole is the root cause.  This is nontrivial because
quadrupoles are closely packed in modern storage rings, the NSLS-II being
no exception.  Therefore, their lattice response vectors (corresponding
columns in $\bs{M}$) are often highly degenerated, especially between
neighboring quadrupoles.  The degeneracy between the $i^{th}$ and $j^{th}$
quadrupole is defined by the correlation coefficient~\cite{Huang:2007zzr}
\begin{equation}\label{eq:corrcoef}
C_{i,j}=\frac{\bs{m}_i\cdot\bs{m}_j}{\Vert\bs{m}_i\Vert\;\Vert\bs{m}_j\Vert},
\end{equation}
where $\bs{m}_i$ is the $i^{th}$ column of $\bs{M}$, which has $N$
elements.  If $|C_{i,j}|$ approaches 1, it is difficult to distinguish
which one is the actual error source with a full Jacobian matrix.  It was
found that rather than using all BPMs, and instead selecting a few
specific BPMs among them, the highly degenerated quadrupoles were
distinguishable.

Consider that there are $N$ BPMs. The $\beta$-beats seen by these BPMs are
$N$-dimension vectors.  Among them, $n\;(n\ll N)$ components can be
selected to form two much shorter sub-vectors $\bs{v}_{i,j}$ in a such way
that $\bs{v}_{i,j}$ have much less correlation between them. This means
they should be as near-orthogonal as possible in an $n$ dimensional vector
space.  There are $N!/(n!(N-n)!)$ different permutations to select
from. We found that it is not difficult to distinguish 5-6 BPMs out of 180
BPMs in the NSLS-II ring even if the correlation between some neighboring
quadrupoles is above 0.98.  Experimentally, we repetitively measure the
lattice functions. Then we compute the correlation coefficients between
$\bs{v}_{i,j}$ and the measured lattice distortion patterns, $\bs{u}$, as
seen only at those 5-6 specific BPMs.  If the quadrupole error was due to
the $i^{th}$ quadrupole, the correlation coefficients
$C_{v_i,u}=\frac{\bs{v}_i\cdot\bs{u}}{\Vert\bs{v}_i\Vert\;\Vert\bs{u}\Vert}$
should be distributed close to $\pm1$. Another one,
$C_{v_j,u}=\frac{\bs{v}_j\cdot\bs{u}}{\Vert\bs{v}_j\Vert\;\Vert\bs{u}\Vert}$
should be around zero, and vice versa.

\begin{figure}[!ht]
\centering \includegraphics[width=1.\columnwidth]{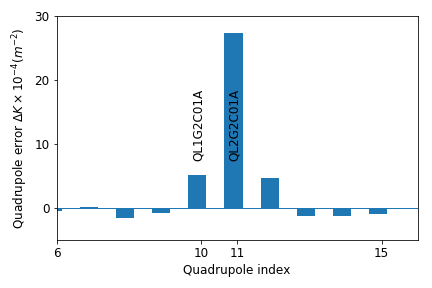}
\caption{\label{fig:q10q11_fitting} The $10^{th}$ quadrupole
  \texttt{QL1G2C01A} excitation is changed by 1 Ampere intentionally at
  the NSLS-II ring.  By observing $\beta_x$ distortion at all 180 BPMs,
  the error source is incorrectly identified as its neighbor, the
  $11^{th}$ quadrupole \texttt{QL2G2C01A}. The reason for that is the
  correlation between the two magnets, as seen by all 180 BPMs, is as high
  as 0.9896.}
\end{figure}

To verify this technique, an experiment and a simulation study were
carried out on the NSLS-II ring.  The excitation current of one quadrupole
\texttt{QL1G2C01A} (with an index of 10) was changed by 1 Ampere.  The
bunch-by-bunch feedback system~\cite{Cheng:2018vfu} was then used to
resonantly drive the beam to perform betatron oscillation at a nearly
constant amplitude.  Beam TbT data was acquired for $800\times1024$ turns.
For every 1024 turns of data, a set of $\beta_x$ functions was extracted
at 180 BPMs.  After averaging them, an error distribution was fitted out
with the Bayesian approach.  It was found that the maximum error is not
\texttt{QL1G2C01A} as it should be, but its neighbor \texttt{QL2G2C01A}
(with an index 11) (see Fig.~\ref{fig:q10q11_fitting}).  It is not
surprising because the correlation between the $10^{th}$ and $11^{th}$
columns of the Jacobian $\bs{M}$ is as high as 0.9896.

\begin{figure}[!ht]
\centering \includegraphics[width=1.\columnwidth]{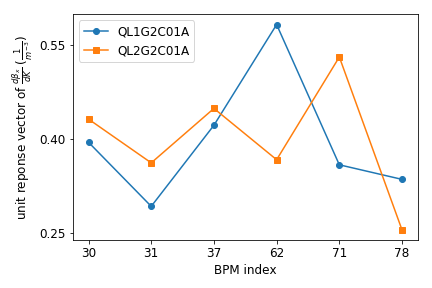}
\caption{\label{fig:q10q11_orth} The unit $\beta_x$ responses vectors of
  quadrupole 10 and 11 seen at 6 selected BPMs. They are near-orthogonal
  in the 6-D vector space because their correlation calculated with
  Eq.~\eqref{eq:corrcoef} is as low as 0.0612.}
\end{figure}

\begin{figure}[!ht]
\centering \includegraphics[width=1.\columnwidth]{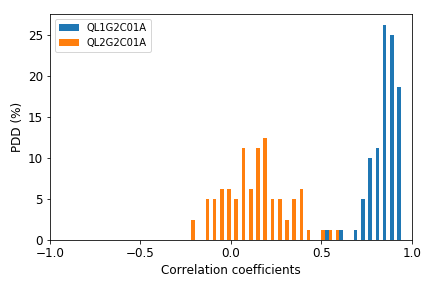}
\caption{\label{fig:q10q11_meas} The probability density distribution
  (PDD) of the correlation coefficients between 800 measured $\beta_x$
  distortion and two unit vectors $\bs{v}_{10,11}$. The independently and
  repeatedly measured $\beta$-beats are highly correlated with quadrupole
  10's pattern, rather than its neighbor.  Based on that we can conclude
  that the actual error source is more likely from the quadrupole 10
  (\texttt{QL1G2C01A}) instead of quadrupole 11 (\texttt{QL2G2C01A}).}
\end{figure}

\begin{figure}[!ht]
\centering \includegraphics[width=1.\columnwidth]{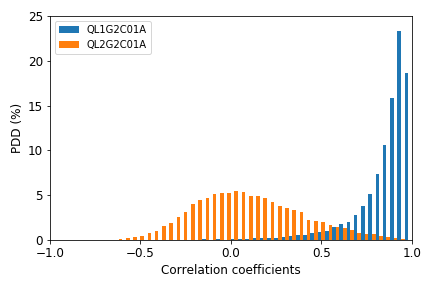}
\caption{\label{fig:q10q11_simu} The probability density distribution
  (PDD) of the correlation coefficients between 25,000 simulated
  $\beta_x$-beats patterns and two unit vectors $\bs{v}_{10,11}$. The
  reproducibility further confirms our experiment observation.}
\end{figure}

Among 180 BPMs, we specifically selected 6 of them with their indices as
\{30, 31, 37, 62, 71, 78\}.  Observed at these BPMs, the unit $\beta_x$
functions response to these two quadrupoles is near-orthogonal with a
correlation coefficient as low as 0.0612 (see Fig.~\ref{fig:q10q11_orth}).
800 independently measured $\beta_x$-beat patterns at those specific 6
BPMs were compared against these two unit response vectors
$\bs{v}_{10,11}$.  The histograms of their correlation coefficients are
illustrated in Fig.~\ref{fig:q10q11_meas}.  It becomes clear that the
$\beta_x$ distortion is likely due to the quadrupole \texttt{QL1G2C01A}
rather than its neighbor \texttt{QL2G2C01A}, because the measured optics
distortion pattern is highly correlated with its response vector.  A
simulation was also performed to reproduce the experiment observation as
illustrated in Fig.~\ref{fig:q10q11_simu}.

\section{\label{sect:summary}summary}
The Bayesian approach explicitly emphasizes the coherence in the existing
methods for linear lattice correction.  By representing the lattice models
as several likelihood functions and some prior probability distributions,
overfitting of the optics corrections can be prevented.  Prior
probabilities can be calculated based on the lattice model prior to
lattice correction.  At the same time, the Bayesian approach gives the
weights for different fitting objectives based on their measurement errors
so that no objectives are over-emphasized by sacrificing others.  If a
distorted $\beta$ pattern comes from a single error source that is highly
degenerated with its neighbors, it can be difficult to address.  A new
technique for resolving degeneracy and identifying the real error source
has been demonstrated with both simulation and experimental observation.

The Bayesian approach is general and can be incorporated into other
lattice and orbit correction algorithms or online
optimizations~\cite{McIntire:2016fnl}, such as the linear optics from
closed orbits (LOCO) algorithm~\cite{Safranek:1997mra}.  The distortions
of orbit response matrix (ORM) elements, due to each individual quadrupole
error, can be calculated in advance and used to explicitly define the
regularization coefficient to control overfitting~\cite{Huang:2007zzr}.
Using a few specific ORM elements to compose orthogonal vectors should be
able to address quadrupole degeneracy as well.  One advantage of LOCO is
that it unifies the $\beta$ and $\phi$ as the directly measurable
parameters with their intrinsic correlation.  Therefore, the fitting needs
to be driven by a complete lattice model, and might be time-consuming.  In
some scenarios, the prolonged time the LOCO method would require may be
considered worth it.  For example, when TbT data is polluted by the beam
decoherence due to a large positive chromaticity and/or
nonlinearity~\cite{Meller:1987ug}.

\begin{acknowledgments}
Y. Li would like to thank Dr. X. Huang (SLAC) and Prof. Y. Hao (Michigan
State Univ.) for some fruitful discussions.  This work was supported by
the U.S. Department of Energy under Contract No. DE-SC0012704.

\end{acknowledgments}
 
\bibliography{probview.bib}

\end{document}